\documentclass[prl,twocolumn,showpacs,groupedaddress]{revtex4}

\usepackage{graphicx}
\usepackage{amssymb}
\usepackage{amsmath}
\usepackage{amsfonts}

\begin{document}

\title{Fault-Tolerant Quantum Computation via Exchange interactions}
\author{M. Mohseni}
\affiliation{Department of Physics, University of Toronto, 60 St. George St., Toronto,
ON, M5S 1A7, Canada}
\author{D.A. Lidar}
\affiliation{Chemical Physical Theory Group, and Center for Quantum Information and
Quantum Control, University of Toronto, 80 St. George St., Toronto, ON, M5S
3H6, Canada}
\pacs{03.67.Lx,03.67.Pp,03.65.Yz}

\begin{abstract}
Quantum computation can be performed by encoding logical qubits
into the states of two or more physical qubits, and controlling a
single effective exchange interaction and possibly a global
magnetic field. This ``encoded universality'' paradigm offers potential
simplifications in quantum computer design since it does away with
the need to perform single-qubit rotations. Here we show that
encoded universality schemes can be combined with quantum error correction. In particular,
we show explicitly how to perform
fault-tolerant leakage correction, thus overcoming the main obstacle to fault-tolerant encoded
universality.
\end{abstract}

\maketitle

In the ``standard paradigm'' of quantum computing (QC) a universal set of
quantum logic gates is enacted via the application of a complete set of
single-qubit gates, along with a non-trivial (entangling) two-qubit gate
\cite{Nielsen:book}. It is in this context that the theory of fault tolerant
quantum error correction (QEC) (e.g., \cite{Gottesman:97bPreskill:99}), and
the well-known associated threshold results (e.g., \cite{Knill:98Steane:03}), have been developed. These results are of crucial importance since they
establish the in-principle viability of QC, despite the adverse effects of
decoherence and inherently inaccurate controls. However, some of the
assumptions underpinning the standard paradigm may translate into severe
technical difficulties in the laboratory implementation of QC, in particular
in solid-state devices. Any quantum system comes equipped with a set of
``naturally available'' interactions, i.e., interactions which are inherent
to the system and are determined by its symmetries, and are most easily
controllable. For example, the symmetries of the Coulomb interaction dictate
the special scalar form of the Heisenberg exchange interaction, which
features in a number of promising solid-state QC proposals \cite{Loss:98Kane:98Vrijen:00}. The introduction of single-spin operations
requires a departure from this symmetry, and typically leads to
complications, such as highly localized magnetic fields \cite{LidarThywissen:04}, powerful microwave radiation that can cause excessive
heating, or $g$-tensor engineering/modulation \cite{Yablonovitch:03}. For
these reasons the ``encoded universality'' (EU) alternative to the standard
paradigm has been developed (e.g., \cite{Zanardi:04}). In EU, single-qubit
interactions with external control fields are replaced by ``encoded''
single-qubit operations, implemented on logical qubits via control of
exchange interactions between their constituent physical qubits. It has been
shown that such an exchange-only approach is also capable of universal QC,
on the (decoherence-free) subspace spanned by the encoded qubits \cite{Bacon:99aKempe:00}. Explicit pulse sequences have been worked out for the
implementation of encoded logic gates in the case when only the exchange
interaction is available \cite{DiVincenzo:00aVala:02,Kempe:01}, which can be
simplified by assuming the controllability of a global, time-dependent
magnetic field \cite{LidarWu:01,Levy:01}.

The issue of the robustness of encoded universal QC in the presence of
decoherence has been addressed in a number of publications, mostly using a
combination of decoherence-free subspaces (DFSs) and dynamical decoupling
methods
\cite{WuByrdLidar:02Viola:01aLidarWu:02Zhang:04,Kempe:01}.
However, in contrast to the case of the standard paradigm, so far a
theory 
of fault tolerant QEC has not yet been developed for encoded universal QC.
The difficulty originates from the fact that EU constructions use only a
subspace of the full system Hilbert space, and hence are subject to leakage
errors to the orthogonal subspace. Standard QEC theory then breaks down
under the restriction of using only a limited set of interactions, since
these interactions are not universal over the orthogonal subspace, and
cannot, using pre-established methods, be used to fix the leakage problem.
Here we show for the first time how to extend the theory of fault tolerant
QEC so as to encompass encoded universal QC. This establishes also the fault
tolerance of a class of DFSs, for which prior fault tolerance results were
of a heuristic nature \cite{Lidar:PRL99}.

\textit{Encoded Universality}.--- We first briefly review the concept of EU
in the context of a particularly simple encoding of one logical qubit into
the states of two neighboring physical qubits:\ $|0_{L}\rangle
_{i}=|0_{2i-1}\rangle \otimes |1_{2i}\rangle $, $|1_{L}\rangle
_{i}=|1_{2i-1}\rangle \otimes |0_{2i}\rangle $, where $|0\rangle $
($|1\rangle $)\ is the $+1$ ($-1$) eigenstate of the Pauli matrix
$\sigma _{z}$. We shall refer to this encoding as a ``two-qubit
universal code'' (2QUC), 
and more generally to EU encodings involving $n$ qubits per logical qubit as
``$n$QUC''. In Ref.~\cite{LidarWu:01} it was shown how to construct a
universal set of encoded quantum logic gates for the 2QUC, generated
from the widely applicable class of 
(effective or real) exchange Hamiltonian of the form $H_{\mathrm{ex}}\equiv
\sum_{i<j}H_{ij}$, where
\begin{equation}
H_{ij}=\sum_{i<j}J_{ij}(X_{i}X_{j}+Y_{i}Y_{j})+J_{ij}^{z}Z_{i}Z_{j}.
\label{eq:Hij}
\end{equation}
Here $X_{i},Y_{i},Z_{i}$ represent the Pauli matrices $\sigma
_{x},\sigma _{y},\sigma _{z}$ acting on the $i$th physical qubit. The Heisenberg interaction
is the case $J_{ij}=J_{ij}^{z}$ (e.g., electron and nuclear spin qubits, \cite{Loss:98Kane:98Vrijen:00}, while the
XXZ and XY models are, respectively, the cases $J_{ij}\neq J_{ij}^{z}\neq 0$
(e.g., electrons on helium, \cite{Platzman:99}) and $J_{ij}\neq
0,J_{ij}^{z}=0$ (e.g., quantum dots in cavities, \cite{Imamoglu:99}). In
essentially all pertinent QC proposals one can control the $J_{ij}$ for $|i-j|\lesssim 2$, though not independently from $J_{ij}^{z}$. As
usual in the EU\ discussion we do not assume that the technically
challenging single-qubit external operations of the form $\sum
f_{i}^{x}(t)X_{i}+f_{i}^{y}(t)Y_{i}$ are available. We do assume
that a (global) free Hamiltonian $H_{0}=\sum_{i}\frac{1}{2}$
$\omega _{i}Z_{i}$ with non-degenerate $\omega _{i}$'s can be
exploited for QC in the sense that the $\omega _{i}$ are
collectively controllable, e.g., via the application of a global
magnetic field.
Note that $\overline{X}_{2i-1,2i} $
and $\overline{Z}_{2i-1,2i}$, where $\overline{X}_{ij}\equiv \frac{1}{2}
(X_{i}X_{i}+Y_{i}Y_{j})$, $\overline{Z}_{ij}\equiv \frac{1}{2}(Z_{i}-Z_{j})$
, generate an su$(2)$ algebra on the $i$th 2QUC, while $\overline{ZZ}
_{i,i+1}\equiv Z_{2i}Z_{2i+1}$ generates a controlled-phase (CP) gate
between the $i,i+1$th 2QUCs. Here bars denote logical operations on the
2QUC, so that, e.g., $\left\vert 0_{L}\right\rangle _{i}\overset{\overline{X}_{2i-1,2i}}{\leftrightarrow }\left\vert 1_{L}\right\rangle _{i}\,$. Given only the
ability to control the $J_{ij}$, explicit encoded logic gates can be derived
using the identity
\begin{eqnarray}
C_{I_{k}}^{\phi }\circ \exp (i\theta I_{i})&\equiv&\exp (-i\phi I_{k})\exp
(i\theta I_{i})\exp (i\phi I_{k})  \notag \\
&=&\exp [i\theta (I_{i}\cos \phi +I_{j}\sin \phi )],  \label{eq:conj}
\end{eqnarray}
valid for su$(2)$ generators satisfying the commutation relations $[I_{i},I_{j}]=iI_{k}$ (and cyclic permutations). E.g., an encoded
CNOT gate over control (subscript $C$, qubits $1,2$) and target
(subscript $T$, qubits
$3,4$)\ 2QUCs can be constructed as follows for the XY model: $\overline{
CNOT}=\overline{W}_{T}\overline{CP}\,\overline{W}_{T}$, where $\overline{W}
_{T}=e^{i\frac{\pi }{2}}e^{-i\frac{\pi
}{4}\overline{X}_{34}}e^{-i\frac{\pi
}{4}\overline{Z}_{34}}e^{-i\frac{\pi }{4}\overline{X}_{34}}$ is
the encoded Hadamard gate,
\begin{equation}
\overline{CP}=i\{C_{\overline{X}_{13}}^{\pi /4}\circ C_{\overline{X}
_{12}}^{\pi /2}\circ e^{-i\frac{\pi
  }{2}\overline{X}_{23}}\}e^{-i\frac{\pi
  }{8}(Z_{1}-Z_{2})}e^{-i\frac{\pi }{8}(Z_{3}-Z_{4})} 
\label{eq:CP-XY}
\end{equation}
is the encoded controlled-phase gate. For the Heisenberg and XXZ
models one has
\begin{equation}
e^{-itJ_{2i,2i+1}^{z}\overline{ZZ}_{i,i+1}}=e^{-itH_{2i,2i+1}}C_{\overline{Z}
_{2i-1,2i}}^{\pi }\circ e^{-itH_{2i,2i+1}},  \label{eq:CP-Heis}
\end{equation}
which is equivalent to the $\overline{CP}$\ gate when $tJ_{2i,2i+1}^{z}=\pi
/4$. Importantly, in all these cases universal encoded QC is possible via
relaxed control assumptions, namely control of only the parameters $J_{i,i+1} $ and a global magnetic field.

\textit{Hybrid 2QUC-Stabilizer codes}.--- Our solution for fault tolerant EU
involves a concatenation of 2QUC and the method of stabilizer codes of QEC
theory \cite{Gottesman:97bPreskill:99}. We define a hybrid $n$QUC-Stabilizer
code (henceforth, ``S-$n$QUC'') as the stabilizer code in which each
physical qubit state $\left\vert \psi \right\rangle = \alpha \left\vert
0\right\rangle + \beta \left\vert 1\right\rangle $ is replaced by the $n$QUC
qubit state $\left\vert \psi _{U}\right\rangle = \alpha \left\vert
0_{U}\right\rangle + \beta \left\vert 1_{U}\right\rangle$. With this
replacement $X_{i}$ on physical qubit $i$ must be replaced by its encoded
version $\overline{X}_{i}$, and similarly for $Y_{i}$ and $Z_{i}$. Thus,
physical-level operations on the stabilizer code are replaced by
encoded-level operations on the 2QUC. This replacement rule also applies to
give the new stabilizer for the S-$n$QUC. For example, suppose we
concatenate the 2QUC with the three-qubit phase-flip code $|+\rangle
^{\otimes 3},|-\rangle ^{\otimes 3}$, where $\left\vert \pm \right\rangle
=(\left\vert 0\right\rangle \pm \left\vert 1\right\rangle )/\sqrt{2}$. The
stabilizer of the latter is generated by $X_{1}X_{2},X_{2}X_{3}$. Then the
stabilizer for the hybrid S-2QUC $\left\vert 0_{H}\right\rangle
=\frac{1}{2\sqrt{2}}(\left\vert 01\right\rangle +\left\vert
10\right\rangle )^{\otimes 
3}$, $\left\vert 1_{H}\right\rangle =\frac{1}{2\sqrt{2}}(\left\vert
01\right\rangle -\left\vert 10\right\rangle )^{\otimes 3}$ is just
$S=\{\overline{X}_{1}\overline{X}_{2},\overline{X}_{2}\overline{X}_{3}\}$,
with $\overline{X}_{i}=X_{2i-1}X_{2i}$.

We will assume that it is possible to make measurements directly in the 2QUC
basis. This involves, e.g., distinguishing a singlet $(|01\rangle
-|10\rangle )/\sqrt{2}$ from a triplet state $(|01\rangle +|10\rangle )/\sqrt{2}$, or performing a non-demolition measurement of the first qubit in
each 2QUC logical qubit; these tasks are currently under active
investigation, e.g., \cite{Friesen:04}. In conjunction with the encoded
universal gate set, it is then evidently possible to perform the entire
repertoire of quantum operations needed to compute fault tolerantly on the
2QUC, using standard stabilizer-QEC methods \cite{Gottesman:97bPreskill:99}.
Note that because the stabilizer code is, in our case, built from 2QUC
qubits, it is \textit{a priori} \emph{not} designed to fix errors on the
physical qubits. Thus, our next task is to consider these physical-level
errors.

\textit{Physical phase flips}.--- Let $\mathcal{C}$ be a stabilizer code
that can correct a single phase flip error, $Z_{i},$ on any of the physical
qubits. Therefore at least one of the generators of its stabilizer
anticommutes with the error $Z_{i}$. This implies that there is at least one
stabilizer generator which includes the operator $X_{i}$ or $Y_{i}$.
Consider the hybrid code $\mathcal{C}^{\prime }$ resulting from
concatenating $\mathcal{C} $ and an $n$QUC. The stabilizer of $\mathcal{C}
^{\prime }$ is found by replacing $X_{i},Y_{i}$ or $Z_{i}$ by $\overline{X}
_{i},\overline{Y}_{i}$ or $\overline{Z}_{i}$ respectively. Therefore at
least one of the generators of the stabilizer of $\mathcal{C}^{\prime }$
includes the operator $\overline{X} _{i}$ or $\overline{Y}_{i}$, for
all $i$. In the case of a 2QUC we have
$\overline{X}_{i}=X_{2i-1}X_{2i}$ and
$\overline{Y}_{i}=Y_{2i-1}Y_{2i}$, both of which anti-commute with
$Z_{2i-1}$ and $Z_{2i}$. Moreover, one readily verifies that arbitrary products of
error operators anti-commute with at least one stabilizer generator, or have
trivial effect. Therefore the corrigibility condition of errors on
stabilizer codes \cite{Nielsen:book} are satisfied, and hence \emph{a phase
flip error on any physical qubit in a hybrid S-2QUC is always correctible}.

\textit{Physical bit flip}.--- In contrast to physical-level phase flips,
bit flips, $\{X_{2i-1},Y_{2i-1},X_{2i},Y_{2i}\}$, cause leakage from the
2QUC subspace via transitions to the orthogonal, ``leakage'' subspace spanned by $\{\left\vert 0_{2i-1}0_{2i}\right\rangle ,\left\vert
1_{2i-1}1_{2i}\right\rangle \}$. The generators of the encoded su$(2)$ on a
2QUC qubit, $\overline{X}_{2i-1,2i},\overline{Z}_{2i-1,2i}$, annihilate this
subspace, and hence will fail to produce the desired effect if used to
implement standard QEC operations.

\textit{Two-physical-qubit errors}.--- Lastly we need to consider the case
of two physical-level errors affecting two qubits of the same 2QUC block
(the case of two errors on two qubits in different 2QUC blocks is already
covered by the considerations above). Listing all possible such errors we
find that (i)
$\{XX=\overline{X},XY=-\overline{Y},YX=\overline{Y},YY=\overline{X},ZZ=-\overline{I}\}$
act as single encoded-qubit errors, and 
thus are correctible by the stabilizer QEC, and (ii) $\{XZ,YZ,ZX,ZY\}$ all
act as leakage errors. We conclude that our task is to find a way to solve
the leakage problem by using only the available interactions. We do this in
two steps: first we construct a ``leakage correction unit'' (LCU) assuming
perfect pulses, then we consider fault tolerance in the presence of 
imperfections in the LCU and computational operations.

\textit{Leakage correction unit}.--- We assume that we can reliably prepare
a 2QUC ancilla qubit in the state $|0_{L}\rangle $. We now define an LCU as
the unitary operator $L$ whose action (up to phase) is:
\begin{eqnarray}
L|0_{L}\rangle |0_{L}\rangle &=&|0_{L}\rangle |0_{L}\rangle \quad
L\left\vert 0_{1}0_{2}\right\rangle |0_{L}\rangle =|0_{L}\rangle \left\vert
0_{3}0_{4}\right\rangle  \notag \\
L|1_{L}\rangle |0_{L}\rangle &=&|1_{L}\rangle |0_{L}\rangle \quad
L\left\vert 1_{1}1_{2}\right\rangle |0_{L}\rangle =|0_{L}\rangle \left\vert
1_{3}1_{4}\right\rangle  \label{eq:LC}
\end{eqnarray}
Here the first (second) qubit is the data (ancilla) qubit, and the action of
$L$ on the remaining $12$ basis states is completely arbitrary. The LCU thus
\emph{conditionally} swaps a leaked data qubit with the ancilla, resetting
the data qubit to $|0_{L}\rangle $; this corresponds to a logical error on
the data qubit, which can be fixed by the stabilizer code. Note that $L$
entangles the data and ancilla qubits, which means that we can determine
with certainty if a leakage correction has occurred or not by measuring the
state of ancilla. We next show how to construct the transformation $L$ from
the available interactions. We decompose $L$ in general as follows: $L=\sqrt{{\sc SWAP}}\times \sqrt{{\sc SWAP}^{\prime }}$, where
\begin{eqnarray}
\sqrt{{\sc SWAP}} &=&\exp [-i\frac{\pi }{4}(\overline{X}_{13}+\overline{X}_{24})]
\label{eq:SWAP} \\
\sqrt{{\sc SWAP}^{\prime }} &=&\exp [-i\frac{\pi }{4}(\overline{X}_{13}Z_{2}Z_{4}+
\overline{X}_{24}Z_{1}Z_{3})]  \label{eq:SWAP'}
\end{eqnarray}
and $\exp [-i\frac{\pi }{4}\overline{X}_{ij}]$ is just the square-root of
swap gate between physical qubits $i$ and $j$. The gate $\sqrt{{\sc SWAP}}$
applies this operation on qubits $1,3$ and $2,4$ in parallel. Depending on
whether the eigenvalues of $Z_{2}Z_{4}$ and $Z_{1}Z_{3}$ are $+1$ or $-1$ on
the four basis states of Eq.~(\ref{eq:LC}), the gates $\sqrt{{\sc SWAP}}$ and $
\sqrt{{\sc SWAP}^{\prime }}$ multiply constructively (destructively) to generate a
full swap (identity).

\begin{figure}[tbp]
  \includegraphics[width=10.3cm]{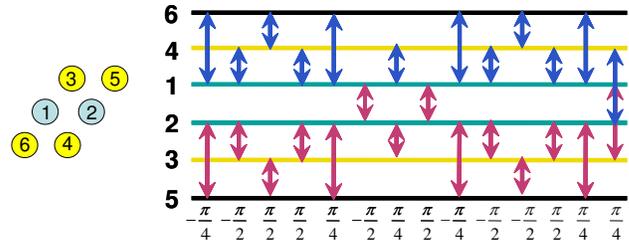}
  \vspace{-4.4cm}
\caption{Circuit for the {$\protect\sqrt{{\sc SWAP}^{\prime}}$}
operation in the
XY model. Time flows from left to right. Data-physical qubits are numbered $1,2$, while $3$-$6$ are ancilla-physical qubits. An angle
$\protect\phi$
under an arrow connecting qubits $i,j$ represents the pulse $\exp[-i\protect\phi\overline{X}_{ij}]$. A total of $13$ such pulses are required.
The circles on the left represent a possible arrangement of qubits
so that all are nearest neighbors throughout the pulse sequence.}
\label{fig1}
\end{figure}

\textit{Circuits for the LCU}.--- Eq.~(\ref{eq:SWAP'}) involves four-body
spin interactions. 
We next show how to construct these from available two-body
interactions. For systems with XY-type of exchange interactions
\cite{Imamoglu:99} the $\sqrt{{\sc SWAP}}$ gate consumes a single pulse.
A circuit for performing $\sqrt{{\sc SWAP}^{\prime }} $ is given in
Fig.~\ref{fig1}.

For the class of Heisenberg systems \cite{Loss:98Kane:98Vrijen:00}, and for
XXZ-type systems \cite{Platzman:99}, we refocus the Ising term $J_{ij}^{z}Z_{i}Z_{j}$, and use the following identity:

\begin{eqnarray}
\sqrt{{\sc SWAP}^{\prime }} &=&\{C_{Z_{2}Z_{3}}^{\pi /4}\circ C_{\overline{X}_{12}}^{\pi /2}\circ \exp [-i\pi \overline{X}_{14}/4]\}\times   \notag \\
&&\{C_{Z_{1}Z_{4}}^{\pi /4}\circ C_{\overline{X}_{12}}^{\pi /2}\circ \exp
[-i\pi \overline{X}_{23}/4]\}  \label{eq:SWAP'-XXZ}
\end{eqnarray}
To generate $\overline{{X}}_{ij}$ and $Z_{i}Z_{j}$ we use [recall Eq.~(\ref{eq:Hij})]
\begin{equation}
e^{-itH_{ij}/2}C_{Z_{i}}^{\pi /2}\circ e^{\pm itH_{ij}/2}=e^{-2itJ_{ij}^{x}
\overline{X}_{ij}}\mbox{{\tiny (+)}}~\mathrm{or}~e^{-itJ_{ij}^{z}Z_{i}Z_{j}}
\mbox{{\tiny (-)}}  \label{eq:refocusZZ}
\end{equation}
which is an example of recoupling \cite{LidarWu:01}. The
$Z_{i}$-pulses required for this can, in turn, be generated as
follows:

\begin{equation}
e^{it\sum_{l}\frac{1}{2}\omega _{l}Z_{l}}C_{H_{ik}}^{\pi /4}\circ
e^{-it\sum_{l}\frac{1}{2}\omega _{l}Z_{l}}=e^{\frac{1}{2}it\Delta
_{ki}Z_{k}}e^{\frac{1}{2}it\Delta _{ik}Z_{i}}
\label{eq:single-Z}
\end{equation}
where $i,k\in \{1,...,6\}$ and $\Delta _{ik}\equiv \omega _{i}-\omega _{k}$.
By adjusting the time so that $t\Delta _{ki}=\pi $ and inserting Eq.~(\ref{eq:single-Z}) into Eq.~(\ref{eq:refocusZZ}) we generate the pulses $\exp
[i\pi Z_{i}/2]$ required in the conjugation step of Eq.~(\ref{eq:refocusZZ}), since the action on qubit $k$ cancels out. Note that all spins not
participating in the exchange interaction are unaffected by the procedure of
Eq.~(\ref{eq:single-Z}).
For all types of exchange interactions we have checked that the $\sqrt{{\sc SWAP}^{\prime }}$
can be also performed using only the two physical ancilla qubits $3,4$, with
the same number of physical pulses, by
sacrificing to some degree the possibility
of parallel operations within each LCU. In all cases the time required for
realizing the LCU is, to within a factor of two, equal to that for
performing a single $\overline{CNOT}$. We note that a
non-unitary QEC leakage detection circuit was described in
Ref.~\cite{Gottesman:97bPreskill:99}. Unfortunately, this method
is not in general applicable to $n$QUCs, since
the required logic gates operate over the full system Hilbert
space. \emph{Constraints} for unitary leakage-correcting operations,
similar to our LCU, 
were derived in Ref.~\cite{Kempe:01} for the 3QUC and Heisenberg-only
computation, but no explicit circuit was given there.

\textit{Fault-tolerant computation on the S-2QUC}.--- So far we have assumed
perfect gates. We now relax this assumption. Fault-tolerant computation is
defined as a procedure in which if any component of a circuit fails to
operate, at most one error appears in each encoded-block qubit \cite{Nielsen:book,Gottesman:97bPreskill:99}. For a specific component to be
fault-tolerant, the probability of error per operation should be below a
certain threshold \cite{Knill:98Steane:03}. \emph{Transversal} quantum
operations, such as the the normalizer elements CNOT, phase, and Hadamard ($W$), are those which can be implemented in pairwise fashion over physical
qubits. This ensures that an error from an encoded block of qubits cannot
spread into more than one physical qubit in another encoded block of qubit
\cite{Nielsen:book,Gottesman:97bPreskill:99}. Transversal operations become
automatically fault-tolerant. In order to construct a universal
fault-tolerant set of gates we should in addition be able to implement,
e.g., a fault-tolerant encoded $\pi /8$ gate; although this gate is not
transversal it can be realized by performing fault-tolerant measurements
\cite{Nielsen:book}. Let us denote by a double bar encoded gates that act on
the S-$n$QUC. It is easy to see that $\overline{\overline{CNOT}}$ and $\overline{\overline{W}}$ can be implemented transversally using EU
operations as above. Moreover, by inspection of Ref.~ \cite{Nielsen:book} it
is easy to see that all operations needed to construct the $\pi /8$ gate, in
particular fault tolerant measurements and cat state preparation, can be
done in the 2QUC basis, without any modification, as long as one can measure
directly in the 2QUC basis (as discussed above). Hence, with respect to
\emph{logical} errors on the 2QUC qubits, the hybrid S-$n$QUC preserves all
the required fault-tolerance properties.

This leaves physical-level phase and bit flip errors during \emph{encoded
logic gates}. We already showed that phase flip errors act as logical errors
that the stabilizer QEC can correct. Bit flip errors are more problematic: a
\emph{single} leakage error invalidates the stabilizer code block in which
it occurs, since the QEC procedures are ineffective in the leakage subspace.
Hence if such errors were to propagate during a logic operation such as $\overline{\overline{CNOT}}$, they would -- if left uncorrected -- overwhelm
the stabilizer level and result in catastrophic failure.
We have verified that leakage errors propagate as either: (i) single
physical-level leakage errors, remaining localized on the \emph{same} qubit,
in the case of an error taking place \emph{before} or \emph{between} the
unitary transformations that make up an encoded logic gate \cite{SingleLeak}; (ii) as two-qubit leakage errors if a single-qubit leakage error happened
\emph{during} the latter transformations. In any case, \emph{the solution is
to invoke the LCU after each logic operation, and before the QEC circuitry}.
The LCU turns a leakage error into a logical error, after which multilevel
concatenated QEC \cite{Nielsen:book,Gottesman:97bPreskill:99} can correct
these errors with arbitrary accuracy. However, uncontrolled leakage error
propagation during QEC \emph{syndrome measurements} must be avoided by
inserting LCUs on each 2QUC after the cat-state preparation and before the
verification step.

The final possibility we must contend with are leakage errors taking place
during {\emph{the operation of the LCU itself}. Such a faulty LCU could
incorrectly change the state of the ancilla qubit in Eq.~(\ref{eq:LC}).
Therefore finding the ancilla in either $|00\rangle$ or $|11\rangle$ is an
inconclusive result. Now let $p_{\mathrm{s}}$ be the probability of success
of the LCU operation in one trail (this depends on accurate gating of the
interaction Hamiltonian, etc.). Let $\omega =\mathrm{Tr}(\rho _{\mathrm{f}}\left\vert 0_{L}\right\rangle \left\langle 0_{L}\right\vert )$ be the
probability of finding the ancilla-qubit in the final state $\left\vert
0_{L}\right\rangle $, where $\rho _{\mathrm{f}}$ represents the final
entangled state of data-qubit and ancilla ($\omega$ critically depends on
the quantum channel error model). The probability, $p_{\mathrm{c}}$, of
achieving \emph{conclusive} and \emph{correct} information about the state
of the data-qubit (being in the logical subspace) is $p_{\mathrm{c}}= (\omega \wedge p_{\mathrm{s}})/\omega$. This is the conditional probability
of LCU success when we already know that the ancilla is in state $|0_L\rangle $. Then $1-p_{\mathrm{c}}$ is the probability of achieving a
\emph{conclusive} but \emph{wrong} result. We can arbitrarily boost the
success probability of the LCU+measurement, $1-(1-p_{\mathrm{c}})^{n}$, to
be higher than some constant $c_{\circ }$, by repeating this procedure until
we obtain $n\geq \log _{1-p_{\mathrm{c}}}(1-c_{\circ })$ consecutive
no-leakage events. }

\textit{Conclusion}.--- We have presented a theory of fault-tolerant QC for
systems governed by XY, XXZ or Heisenberg exchange interactions, operated
without single-qubit gates. In doing so, the theories of QEC and EU were
reconciled for the first time by introducing a type of hybrid EU-stabilizer
code. Leakage out of the EU code space was identified as the key problem and
solved here using a fully constructive approach, within the EU framework of
utilizing only the system's intrinsic interactions. Many elements of this
theory can be directly generalized to other quantum systems with a known set
of experimentally available Hamiltonians.
These results confirm the viability of the EU paradigm,
with its associated advantages of reduced quantum control constraints and
improved experimental compatibility to the interactions that are naturally
available in a given quantum system. Moreover, by constructing error
correction operations from a Hamiltonian formulation, rather than from
gates as the elementary
building blocks, a more accurate and reliable calculation of the
fault-tolerance threshold is
possible than in previous approaches. This will be undertaken in a
future publication.

Support from OGSST and NSERC (to M.M.), the DARPA-QuIST program (managed by AFOSR
under agreement No. F49620-01-1-0468) and the Sloan foundation (to D.A.L.),
is gratefully acknowledged. We thank K. Khodjasteh and A. Shabani for
useful discussions.


\begin{thebibliography}{99}
\bibitem{Nielsen:book} {M.A. Nielsen, I.L. Chuang}, \emph{Quantum
Computation and Quantum Information} ({Cambridge University Press},
Cambridge, UK, 2000).

\bibitem{Gottesman:97bPreskill:99} D. Gottesman, eprint
  quant-ph/9705052; {J. Preskill}, in \emph{Introduction to Quantum
    Computation and Information} ({World Scientific}, Singapore,
  1999), edited by H.K. Lo, S. Popescu and T.P. Spiller.

\bibitem{Knill:98Steane:03} {E. Knill, R. Laflamme, W. Zurek}, Science
\textbf{279}, 342 (1998); {A.M. Steane}, Phys. Rev. A \textbf{68}, 042322
(2003).

\bibitem{Loss:98Kane:98Vrijen:00} {D. Loss, D.P. DiVincenzo}, Phys. Rev. A
\textbf{57}, 120 (1998); {B.E. Kane}, Nature \textbf{393}, 133 (1998); {R.
Vrijen \textit{et. al}}, Phys. Rev. A \textbf{62}, 012306 (2000).

\bibitem{LidarThywissen:04} {D.A. Lidar, J.H. Thywissen}, J. Appl. Phys.
\textbf{96}, 754 (2004).

\bibitem{Yablonovitch:03} {E. Yablonovitch \textit{et. al}}, {Proc. of the
IEEE} \textbf{91}, 761 (2003).

\bibitem{Zanardi:04} {P. Zanardi, S. Lloyd}, Phys. Rev. A \textbf{69},
022313 (2004).

\bibitem{Bacon:99aKempe:00} {D. Bacon \textit{et. al}}, Phys. Rev. Lett.
\textbf{85}, 1758 (2000); {J. Kempe \textit{et. al}}, Phys. Rev. A \textbf{63}, 042307 (2001).

\bibitem{DiVincenzo:00aVala:02} {D.P. DiVincenzo \textit{et. al}}, Nature
\textbf{408}, 339 (2000); {J. Vala, K.B. Whaley}, Phys. Rev. A \textbf{66},
022304 (2002).

\bibitem{Kempe:01} {J. Kempe \textit{et. al}}, Quant. Inf. Comp. \textbf{1},
33 (2001).

\bibitem{LidarWu:01} {D.A. Lidar, L.-A. Wu}, Phys. Rev. Lett. \textbf{88},
017905 (2002).

\bibitem{Levy:01} {J. Levy}, Phys. Rev. Lett. \textbf{89}, 147902 (2002).

\bibitem{WuByrdLidar:02Viola:01aLidarWu:02Zhang:04}  {L.-A. Wu, M.S. Byrd, D.A. Lidar},
Phys. Rev. Lett. \textbf{89}, 127901 (2002); {L. Viola},
Phys. Rev. A \textbf{66}, 012307 (2002); {D.A. Lidar, L.-A Wu}, Phys.
Rev. A \textbf{67}, 032313 (2003); {Y. Zhang \textit{et. al}}, Phys. Rev. A
\textbf{69}, 042315 (2004).

\bibitem{Lidar:PRL99} {D.A. Lidar, D. Bacon, K.B. Whaley}, Phys. Rev. Lett.
\textbf{82}, 4556 (1999).

\bibitem{Platzman:99} {P.M. Platzman, M.I. Dykman}, Science \textbf{284}, 1967 (1999).

\bibitem{Imamoglu:99} {A. Imamo$\bar{\mathrm{g}}$lu \textit{et. al}}, Phys.
Rev. Lett. \textbf{83}, 4204 (1999).

\bibitem{Friesen:04} {M. Friesen \textit{et. al}}, Phys. Rev. Lett. \textbf{92}, 037901 (2004).

\bibitem{SingleLeak} To see this consider a generic unitary transformation $
G_{ij}\in \{H_{ij}=J_{ij}(X_{i}X_{j}+Y_{i}Y_{j})+J_{ij}^{z}Z_{i}Z_{j},
\overline{Z}_{ij}=(Z_{i}-Z_{j})/2\}$, and a single qubit errors $E_{i}\in
\{X_{i},Z_{i}\}$. Then, using $U\exp (A)U^{\dag }=\exp (UAU^{\dag })$ for
unitary $U$ we can commute $E_{i}$ to the left while flipping signs in $
G_{ij}$ appropriately [e.g., $H_{ij}X_{i}=X_{i}
\{J_{ij}(X_{i}X_{j}-Y_{i}Y_{j})-J_{ij}^{z}Z_{i}Z_{j}\}$]. The
transformations with flipped sign combine to give a faulty logic gate on the
2QUC qubits, which is followed by the same $E_{i}$ error.
\end{thebibliography}
\end{document}